\documentstyle[12pt]{article}
\newcommand{\bce}{\begin{center}}
\newcommand{\ece}{\end{center}}
\newcommand{\beq}{\begin{equation}}
\newcommand{\eeq}{\end{equation}}
\newcommand{\bea}{\vspace{0.25cm}\begin{eqnarray}}
\newcommand{\eea}{\end{eqnarray}}

\newcommand{\ba}{\begin{array}}
\newcommand{\ea}{\end{array}}

\newcommand{\doublespace}{
    \renewcommand{\baselinestretch}{1.6}\large\normalsize}

\def\lsim{\mathrel{\rlap{\lower4pt\hbox{\hskip1pt$\sim$}}
    \raise1pt\hbox{$<$}}}     %less than or approx. symbol
\def\gsim{\mathrel{\rlap{\lower4pt\hbox{\hskip1pt$\sim$}}
    \raise1pt\hbox{$>$}}}     %greater than or approx. symbol

\def\lsim{\mathrel{\rlap{\lower4pt\hbox{\hskip1pt$\sim$}}
    \raise1pt\hbox{$<$}}}         %less than or approx. symbol
\def\gsim{\mathrel{\rlap{\lower4pt\hbox{\hskip1pt$\sim$}}
    \raise1pt\hbox{$>$}}}         %greater than or approx. symbol

\def\beq{\begin{equation}}
\def\endeq{\end{equation}}
\def\arr{\begin{eqnarray}}
\def\endarr{\end{eqnarray}}

\makeindex
\begin{document}
%\large
\phantom{.}\hspace{8.5cm}{\Large \bf KFA-IKP(Th)-1994-34} \\
\phantom{.}\hspace{9.5cm}{\large \bf 15 September 1994}
\vspace{2cm}
\begin{center}
{\bf \huge Final state interaction
effects in $D(e,e'p)$ scattering \\}
\vspace{1cm}
{\bf A.Bianconi$^{1,2)}$, S.Jeschonnek$^{3)}$,
N.N.Nikolaev$^{3,4)}$, B.G.Zakharov$^{3,4}$ } \medskip\\
{\small \sl
$^{1)}$Istituto Nazionale di Fisica Nucleare,
Sezione di Pavia, Pavia, Italy \\
$^{2)}$Dipartimento Fisica Nucleare e Teorica,
Universit\`a di Pavia, Italy \\
$^{3)}$IKP(Theorie), Forschungszentrum  J\"ulich GmbH.,\\
D-52425 J\"ulich, Germany \\
$^{4)}$L.D.Landau Institute for Theoretical Physics, \\
GSP-1, 117940, ul.Kosygina 2, V-334 Moscow, Russia
\vspace{1cm}\\}
{\bf \LARGE A b s t r a c t \bigskip\\}
\end{center}
We present a systematic study of the final-state
interaction (FSI) effects in $D(e,e'p)$ scattering in the CEBAF
energy range with particular emphasis on the phenomenon of the
angular anisotropy of the missing momentum distribution. We
find that FSI effects dominate at missing momentum $p_m \gsim
1.5 $fm$^{-1}$. FSI effects in the excitation of the $S$-wave
state are much stronger than in the excitation of the $D$-wave.

\medskip
%{\bf PACS: 25.30Fj,~24.10Eq}

%--------------------------------------------------
\newpage
\doublespace

There is a basic necessity of the knowledge of the nucleon
momentum distribution $n(\vec{p})$ in nuclei, in particular at large
momenta, for it provides unique information on short-distance
nucleon-nucleon interaction (initial-state $N$-$N$ correlations ISC)
in the nuclear medium ([1], for a recent review see [2]). A
quantitative understanding of the final-state
interaction (FSI) effects is crucial for disentangling $n(\vec{p})$
from the missing momentum distribution $W(\vec{p}_{m})$
as measured in inclusive $A(e,e'p)$ scattering in
quasielastic kinematics. In a previous paper [3]
we presented a classification of FSI effects and a comparison
of the relative significance of FSI and short-distance correlation
effects in $^{4}He(e,e'p)$ scattering. Our major finding was that
at large missing momenta $\vec{p}_{m}$, which are usually believed
to come from the ISC effects, the FSI effects are large and
overwhelm the ISC effects (for the related discussion of
FSI effects in heavy nuclei
see also [4-6]). In [3] the FSI was shown to produce
a strong angular anisotropy of the missing momentum distribution
$W(\vec{p}_{m})$, which at large $p_{m}$ develops a sharp peak
in transverse kinematics, at $\theta$ $\sim$ $90^o$, two
minor peaks at $\theta $ = 0$^o$ and 180$^o$ in parallel
kinematics (here $\theta$ is
the angle between the missing momentum and the $(e,e')$ momentum
transfer $\vec{q}$) and a sizable backward-forward asymmetry.

To generalize these results and get a deeper insight into them,
in this
paper a systematic study of the $D(e,e'p)n$ unpolarized scattering
is performed. Prediction of FSI effects in this process is of
great interest on its own sake, as experiments with the deuterium
target constitute an important part of the experimental
program at CEBAF [7]. Other motivations are that (i) in the
region of dominance of ISC effects, the momentum distribution for
heavy nuclei are expected to closely resemble that in the deuteron [8];
(ii) the realistic models of the deuteron [9] allow an accurate
evaluation of FSI effects in the longitudinal missing momentum
distribution which is important for the $y$-scaling analysis [10];
(iii) one can assess differences between FSI effects for the $S$ and
$D$ waves in the initial $N$-$N$ pair. Despite the deuteron being a
dilute target, we find large FSI effects at $p_{m} \gsim 1.5
$\,fm$^{-1}$.

One of the points we wish to address in more detail in this paper
are forward and backward peaks in $W(\vec{p}_{m})$, which have their
origin in the idealized step-function factor $\theta(z)$ in the
eikonal Green function derived [11] under an implicit assumption of
pointlike nucleons ($z$ is the longitudinal separation of nucleons).
We analyze the sensitivity of the forward and backward peaks to the
smearing of the $\theta$-function to account for the finite size of
nucleons and conclude that predictions at $p_{m} \lsim 3$fm$^{-1}$ are
free of uncertainties with the smearing.

We wish to focus on FSI effects, and for the sake of simplicity
we consider the photon as a scalar operator (for the calculation of
the full hadronic tensor in deuteron scattering, see ref.[12]).
Then, the reduced nuclear amplitude for the exclusive process
$D(e,e'p)n$ is given by ${\cal M}_{ji}=\int d^{3}\vec{r}
\exp(i\vec{p}_{m}\vec{r}\,)\chi_{j}^{*}S(\vec{r}\,)\Psi_i(\vec{r}\,)$,
where $\vec r$ $\equiv$ $\vec r_p - \vec r_n$,~
$\Psi_i(\vec{r})$ is the deuteron ground state wave function
for the spin state $i$ and $\chi_{j}$ stands for the spin wave
function of the final $pn$ state. The struck proton is detected
with momentum $\vec{P}$,~ $\vec{q}$ is the $(e,e')$ momentum
transfer and $\vec{p}_m$ $\equiv$ $\vec q - \vec P$ is the missing
momentum. $S(\vec{r})$ is the $S$-matrix of FSI between
struck proton and spectator neutron.
For unpolarized deuterons and for a spin-independent
proton-neutron scattering amplitude, summing over all spin states
of the final state proton and neutron, one finds the observed
momentum distribution
\arr
W(\vec p_m)\ =\ {1\over 3} \sum_{j,i} \vert M_{ji}\vert^2 =
{1 \over 4\pi(2\pi)^{3}}
\int d^{3}\vec{r} d^{3}\vec{r}\,'\exp[i\vec p_m \cdot
(\vec{r}\,'-\vec r\,)]
S(\vec r) S^{\dagger}(\vec r\,')  \nonumber\\
\Bigg[ {u(r) \over r} {u(r') \over r'}\ +\
{1 \over 2} {w(r) \over r} {w(r') \over r'}
\Bigg(3{(\vec r\cdot \vec r\,')^2 \over (rr')^2} - 1\Bigg)\Bigg],
{}~~~~~~~~~~~
\label{eq:1}
\endarr
where $u/r$ and $w/r$ are the radial wave functions of the S and D
components respectively of the deuteron ground state, with
the normalization $\int dr (u^2+w^2) = 1$. In our calculations
we have used the realistic Bonn wave functions $u(r)$ and $w(r)$ as
parametrized in [9].

At the large $Q^{2}\gsim$ (1-2)\,GeV$^{2}$ of the interest in the
CEBAF experiments, the kinetic energy of the struck proton
$T_{kin}\approx Q^{2}/2m_{p}$ is high and FSI can be described by
Glauber theory [11]. Defining transverse and longitudinal
components $\vec{r}$ $\equiv$ $(\vec{b}+z\hat q)$ we can write
\beq
S(\vec{r}) = 1-\theta(z)\Gamma(\vec{b}),
\label{eq:2}
\endeq
where $\Gamma(\vec{b})$ is the profile function of the proton-neutron
scattering, which at high energy can conveniently be parameterized
as
\beq
\Gamma(\vec{b}) \ \equiv\
{ \sigma_{tot} (1 - i \rho) \over 4 \pi b_{o}^2  }
\exp \Big[-{\vec{b}^2 \over 2 b_{o}^2} \Big]
\label{eq:3}
\endeq
($\rho$ is the ratio of the real to imaginary part of the
forward elastic scattering amplitude). The step-function $\theta(z)$
in (3) tells that
the FSI vanishes unless the spectator neutron was in the
forward hemisphere with respect to the struck proton,
which is an idealization in the world of nucleons which have a
finite size. The Glauber formalism describes quite well
nucleon-nucleus scattering at $T_{kin}\gsim 500$MeV (for a review
see [13]). Here
we present numerical results for $T_{kin} \sim 1$GeV
($Q^{2}\sim 2$\,GeV$^{2}$), when $b_{o}\approx 0.5$fm, $\sigma_{tot}
\approx 40$mb and $\rho\approx -0.4$ [13-15].

Let us start with generic observations on the symmetry properties of
$W(\vec{p}_{m})$. Decompose $\vec{p}_{m}$ into the transverse and
longitudinal components $\vec{p}_{m}=\vec{p}_{\perp}+p_{z}\hat{q}$.
In the plane-wave-impulse approximation (PWIA) $S(\vec{r})=1$
and (\ref{eq:1}) gives the isotropic distribution
$W(p_{m})=n(p_{m})$. The radius $b_{0}$ of FSI is
much smaller than the radius of the deuteron $R_{D} \sim 2$fm.
Consequently, as compared to $u(r), w(r)$, the FSI operator
$\theta(z)\Gamma(\vec{b})$ is a "short-ranged" function of
$\vec{b}$ and a "long-ranged" function of $z$, and
this anisotropy of $S(\vec r\,)S^{\dagger}(\vec r\,')$ leads
to the angular anisotropy of $W(\vec{p}_{m})$ shown in Figs.~1,2.
One of the striking effects is the forward-backward asymmetry
$W(p_{\perp},-p_{z})\neq W(p_{\perp},p_{z})$, which is clearly
seen in Fig.~1,2 (for the discussion of this effect in
heavy nuclei see [4,6]). Its origin is in
the nonvanishing real part of the $p$-$n$ scattering amplitude
$\rho\neq 0$, which makes
$S(b,z)S^{\dagger}(b\,',z') \neq S(b',z')S^{\dagger}(b,z)$
in the integrand of (\ref{eq:1}).
At small $p_{m} \ll 1$\,fm$^{-1}$ the angular
distribution is isotropic, with increasing $p_{m}$ it first
develops an approximately symmetric dip at $\theta =90^{o}$. Fig.~2
shows how this dip evolves, through the very asymmetric stage, into
the approximately symmetric peak at $90^{o}$ at larger values of
$p_{m}$. Fig.~1d shows the decomposition of the same $90^{o}$
distribution into PWIA and FSI components. Shown by the
dotted curve is the PWIA distribution, the dashed curve shows the
pure rescattering contribution $\propto \Gamma^{*}\Gamma$ in
Eq.~(\ref{eq:1}). The solid curve includes also the interference
of the FSI and PWIA amplitudes, the size of this
interference term  $\propto(\Gamma +\Gamma^{*})$
(not shown separately here) is given by the
difference between the solid curve and the sum of the dashed and
dotted curves and is most significant around $p_{m}\sim 1.5$\,fm$
^{-1}$. The sharp peak at $\theta = 90^{o}$, which emerges at $p_{m}
\gsim 1.7$\,fm$^{-1}$, is completely dominated by the FSI (elastic
$p$-$n$ rescattering) effect.

It is well known ([8] and references therein) that in the PWIA,
the large-$p_{m}$ tail of the momentum distribution $n(p_{m})$
is dominated by the $D$-wave contribution. The fact that
$\Gamma(b)$ is a "short-ranged" function, whereas the $D$-wave
function $w(r)$ is strongly suppressed at small distances by the
centrifugal barrier, implies that FSI effects in the $D$-wave
contribution to $W(p_{m})$ should be much weaker than in the
$S$-wave contribution. Indeed, Fig.~2 shows that although
the $D$-wave contribution to $W(\vec{p}_{m})$ develops
a nontrivial angular dependence, the overall departure from
the isotropic PWIA distribution is rather small. Fig.~1(a,b) shows
that the $D$-wave contribution remains a dominant component of
$W(\vec{p}_{m})$ in parallel kinematics, but not in
transverse kinematics, see Fig.~1(c,d).

The angular asymmetry mostly comes from the $S$-wave contribution
which can conveniently be written as $W_{S}(\vec{p}_{m})=2^{-5}
\pi^{-4}U^{2}(\vec{p}_{m})$, where $U(\vec{p}_{m})$ can be decomposed
into the PWIA and FSI terms according to the expansion (\ref{eq:2})
\beq
U(\vec{p}_m) =
\int d^{3}\vec{r}\exp(-i\vec{p}_m \cdot \vec r)
S(\vec r) {u(r) \over r}=
u(1;\vec{p}_{m})-u(\Gamma;\vec{p}_{m})
\label{eq:4}
\endeq
The PWIA term $u(1;\vec{p}_{m})$ in (\ref{eq:4}) decreases on the
scale $p_{\perp},p_{z}\sim 1/R_{D}$. In view of $b_{0}^{2} \ll
R_{D}^{2}$, the FSI term $u(\Gamma;\vec{p}_{m})$ has the $p_{\perp}$
dependence $\sim \exp(-{1\over 2}b_{0}^{2}p_{\perp}^{2})$ and at
moderately large $p_{z}$ it decreases on the scale $p_{z} \sim
1/R_{D}$ [3,4]. It has the small absolute normalization $\sim\sigma
_{tot }(pn)/(2\pi R_{D}^{2})$ [3,4]. The destructive interference
of the PWIA and FSI amplitudes produces a dip in the $S$-wave
contribution at $p_{\perp} \sim 1.3
$\,fm$^{-1}$ as shown in Fig.~1c, which is partly filled by the effect
of the finite $\rho$.

Fig.~1b shows that for $p_{m} \gsim 1.5$\,fm$^{-1}$, FSI effects are
substantial also in parallel kinematics. In parallel
kinematics, the most interesting
effect first noticed in [3], is the emergence of forward and
backward peaks in $W(p_{m})$ with increasing $p_{m}$ as shown in
Fig.~3. In Fig.~1b, the same effect shows up as the FSI component
overwhelming the PWIA component at $p_{m} \gsim 3$\,fm$^{-1}$.
The origin of this phenomenon is in the high-frequency Fourier
components of the step-function $\theta(z)$ in the integrands of
(\ref{eq:1},\ref{eq:4}). The dominant effect comes from the "elastic
rescattering" operator $\theta(z)\theta(z')\Gamma(b)\Gamma^{*}(b')$
in the integrand of (\ref{eq:1}), and Fig.~1b shows that the excess
over the PWIA curve at $p_{m}\gsim 3$\,fm$^{-1}$ is entirely due
to the contribution from the term $\propto \Gamma^{*}\Gamma$.
The
scale for the intranucleon separation at which nucleons can still
be treated as approximately structureless particles interacting
with the free-nucleon amplitude, is set by the radius of the
nucleon and/or the
short-range correlation radius $r_{c} \sim 0.5$\,fm. The effect of
the non-pointlike nucleons can be modelled substituting the
idealized step-function $\theta(z)$ for the smeared one
\arr
T(z)\ \equiv\
{1 \over 2} [1+\tanh({z \over z_o})].
\label{eq:5}
\endarr
The educated guess is $z_{0} \lsim {1\over 2}r_{c}$. In Fig.~3 we
show the effect of smearing for $z_{0} = {1\over 2}r_{c}=0.25$\,fm.
We conclude that the uncertainties with the smearing are small at
least up to $p_{m} \lsim 3$\,fm$^{-1}$. Description of the region
of higher $p_{m}$ requires new approaches which go beyond the
Glauber model. We only wish to emphasize that this effect persists
at high $Q^{2}$.

The forward-backward asymmetry in parallel
kinematics is displayed also in Fig.~4, in which we show
the $p_{m}$-dependence of
$$
A_{FB}={W(\theta=0^{o},p_{m})-W(\theta=180^{o},p_{m}) \over
 W(\theta=0^{o},p_{m})+W(\theta=180^{o},p_{m})} \,. $$
 The FSI produces a
large asymmetry $A_{FB}$,
which should be taken into account when looking
for the forward-backward asymmetry which is expected at higher
$Q^{2}$ because of the colour transparency effect [16] (see also
a discussion in [6]).

Still another interesting quantity is the $p_{\perp}$-integrated,
longitudinal missing momentum distribution $F(p_{z})\ \equiv\
\int d^2 \vec{p}_{\perp} W(\vec{p}_{\perp},p_{z})$. In the PWIA,
the distribution
$F_{PWIA}(p_{z})=\int d^{2}\vec{p}_{\perp}n(p_{\perp},y)$
is related to the asymptotic $y$-scaling function [10].
In Fig.~5 we show the ratio $R(p_{z})=F(p_{z})/F_{PWIA}(p_{z})$.
At small $p_{z}$, this ratio
exhibits a depletion $\sim 7$\% in agreement with the estimate
[4] and the NE18 experimental finding [17]. This depletion comes
from the reduction of the flux of protons for the absorption by
the neutron. At larger missing momenta $p_{z}\gsim 1.2$\,fm$^{-1}$,
the contribution $\propto \Gamma\Gamma^{*}$
from FSI enhances $F(p _{z})$
by $\sim $(30-40)\% as compared to the PWIA distribution
$F_{PWIA}(p_{z})$. At large $Q^{2}\gsim 2$\,GeV$^{2}$ and high kinetic
energy of the struck proton $T_{kin}=Q^{2}/2m_{p}\gsim 1$\,GeV, the
proton-neutron total and elastic cross sections are approximately
constant [15]. Therefore, this large FSI effect in $F(p_{z})$ should
persist at large $Q^{2}\gg 1$\,GeV$^{2}$ and should be
taken into account in the asymptotic $y$-scaling regime, too.
The departure from the PWIA further increases at $p_{z} \gsim
3$\,fm$^{-1}$, but the quantitative description of FSI in this
region of $p_{m}$ requires improving upon the Glauber approach,
which goes beyond the scope of this paper.

To summarize the main points, in this analysis of $D(e,e'p)$
scattering in the few-GeV region we have found that FSI effects
completely dominate the observed missing momentum distribution
$W(\vec{p}_{m})$ at high $p_m$ (over $\sim$1.5 $fm^{-1}$). They
produce a marked angular anisotropy in $W(\vec{p}_{m})$,
consisting in a dip at $\theta$ = 90$^o$ (in transverse
kinematics) for $p_m \sim$(1-1.5)\,fm$^{-1}$, and a prominent
peak at the same angle for larger $p_m$. We have shown that
this anisotropy is mostly due to the FSI distortions of the
$S$-wave contribution. The anisotropy effects in the $D$-wave
contribution are also sizable, but much smaller than in the
$S$-wave contribution. The $D$-wave
contribution remains the dominant component for $p_{m}\gsim
1.5$\,fm$^{-1}$ in parallel kinematics. The forward and
backward peaks in the angular distribution at large $p_{m}
\gsim 3$\,fm$^{-1}$ are sensitive to modifications of the
idealized $\theta$-function in the eikonal Green function
and the FSI operator (\ref{eq:2}).
As far as the applicability of Glauber's multiple scattering
theory is concerned, this is an entirely new situation and
shows that the Glauber treatment of FSI in $A(e,e'p)$ scattering
is not fully self-contained at large longitudinal missing momenta,
in contrast to a very accurate
description of the proton-nucleus elastic scattering [13].
The major difference is
that in the proton-nucleus scattering, one has the well defined
asymptotic $|in\rangle$ and $|out\rangle$ proton states, whereas
in FSI in the $D(e,e'p)$ scattering the incoming proton wave
is generated at a finite distance from the target neutron. When
this distance becomes of the order of the radius of nucleons
and/or the radius of $N$-$N$ correlations, one can not describe
the distortion of the wave function of the spectator neutron by
formula (\ref{eq:2}) with the idealized step-function.

We considered the charge operator relevant to the longitudinal
response, but evidently our principle conclusions on FSI are
also applicable to the transverse response.

We found large, $\sim 40$\%, FSI corrections to the longitudinal
missing momentum distribution.
FSI in the $D(e,e'p)$ scattering was discussed also in the recent
work [18] using a different technique. These authors focused on
corrections to the $y$-scaling analysis and did not
consider the angular anisotropy of FSI effects. The numerical
estimates of [18] for the FSI corrections to the longitudinal
missing momentum distribution are close to ours.

{\bf Acknowledgments:} One of us (A.B.) acknowledges previous
discussions with S.Boffi on the general treatment of the
photon-deuteron interaction. This work has been done during a
visit of A.B. at IKP, KFA J\"ulich (Germany), supported by
IKP and by INFN (Italy). A.B. thanks J.Speth for the hospitality
at IKP. This Germany-Italy exchange program was supported in part
by the Vigoni Program of DAAD (Germany) and of the Conferenza
Permanente dei Rettori (Italy). This work was also supported by
the INTAS Grant No. 93-239.
\pagebreak\\
% = = = = = = = = = = = = = = = = = = = = = = = = = = = = = = = = = = =

\pagebreak

{\bf \Large Figure captions:}

\begin{itemize}

\item[{\bf Fig.~1}]
   The missing momentum distribution $W(\vec{p}_{m})$ {\sl vs.}
$p_{m}$ and
its decomposition into {\sl (a,c)} the $S$ and $D$-wave
contributions and {\sl (b,d)} into the PWIA and the
rescattering contributions
$\propto \Gamma^{*}\Gamma$
at {\sl (a,b)}
$\theta=0^{o}$ and {\sl (c,d)} $\theta=90^{o}$. In panels
{\sl (b,d)} the full distribution $W(\vec{p}_{m})$ (solid curve)
includes also the contribution $\propto (\Gamma^{*}+\Gamma)$
from the interference of
the PWIA and FSI amplitudes, which is not shown separately.

\item[{\bf Fig.~2}]
   The angular dependence of (solid curve) the missing momentum
distribution $W(\vec{p}_{m})$
and of its (dotted curve) $S$-wave and (dashed curve) $D$-wave
components at different values of the missing momentum $p_{m}$.

\item[{\bf Fig.~3}]
   The angular dependence of (solid curve) the missing momentum
distribution $W(\vec{p}_{m})$ showing the development of the
forward and backward peaks at large $p_{m}$. The dashed curve
shows $W(\vec{p}_{m})$ found with the smeared step-function
$T(z)$ of Eq.~(\ref{eq:5}) with $z_{0}=0.25$\,fm.

\item[{\bf Fig.~4}]
   The $p_{m}$ dependence of the forward-backward asymmetry
$A_{FB}$ in parallel kinematics.

\item[{\bf Fig.~5}]
   {\sl (a)} The longitudinal missing momentum distribution
$F(p_{z})$ calculated with full FSI (solid curve) and in the
PWIA (dashed curve). {\sl (b)} The ratio of the
full $F(p_{z})$ to the PWIA distribution $F_{PWIA}(p_{z})$.

\end{itemize}

\begin{thebibliography}{299}
%wwwwwwwww
\bibitem{1}
K.Gottfried, {\sl Ann. Phys. (USA)} {\bf 21} (1963) 29; W.Czyz
and K.Gottfried, {\sl Nucl. Phys.} {\bf 21} (1961) 676;
{\sl Ann. Phys. (USA)} {\bf 21} (1963) 47.

\bibitem{2} %%%%
A.E.I.Dieperink and P.K.A. de Witt Huberts,
{\sl Annu. Rev. Nucl. Part. Sci.} {\bf 40} (1990) 239;
S.Boffi, C.Giusti, and F.D.Pacati,
{\sl Phys.Rep.} {\bf 226} (1993) 1.

\bibitem{3} %%%%
A.Bianconi, S.Jeschonnek, N.N.Nikolaev, and B.G.Zakharov,
J\"ulich preprint {\bf KFA-IKP(Th)-1994-29} (1994), to appear
in {\sl Phys. Lett.} {\bf B}.

\bibitem{4} %%%%
N.N.Nikolaev, A.Szczurek, J.Speth, J.Wambach, B.G.Zakharov, and
V.R.Zoller, {\sl Phys. Rev.} {\bf C50} n.3 (1994) R1.

\bibitem{5} %%%%
N.N.Nikolaev, A.Szczurek, J.Speth, J.Wambach, B.G.Zakharov, and
V.R.Zoller, J\"ulich preprint {\bf KFA-IKP(Th)-1993-31} (1994),
submitted to
{\sl Nucl. Phys.} {\bf A}.

\bibitem{6} %%%%
J.Nemchik, N.N.Nikolaev, and B.G.Zakharov, Color transparency after
the NE18 and E665 experiments: Outlook and perspectives at CEBAF,
{\sl J\"ulich preprint} {\bf KFA-IKP(Th)-1994-29} (1994), to be
published in Proceedings of the Workshop on CEBAF at Higher
Energies, 14-16 April 1994.

\bibitem{7} %%%%
J.Mougey (spokesperson), CEBAF Proposal No. E-89-044;
R.G.Milner (spokesperson), CEBAF Proposal No. E-91-007;
D.F.Geesaman (spokesperson), CEBAF Proposal No. E-91-011.

\bibitem{8} %%%%%
C.Marchand et al., {\sl Phys. Rev. Lett.} {\bf 60} (1988) 1703.

\bibitem{9} %%%%
R.Machleidt, K.Holinde, and C.Elster, {\sl Phys. Rep.}
{\bf 149} (1987), 1.

\bibitem{10} %%%%
G.B.West, {\sl Phys. Rep.} {\bf 18}, (1975), 264.
C.Ciofi degli Atti, E.Pace, and G.Salme, {\sl Phys. Rev.} {\bf C43}
(1991), 1155, and references therein.

\bibitem{11} %%%%%
R.J.Glauber, in: {\sl Lectures in Theoretical Physics}, v.1,
ed. W.Brittain and L.G.Dunham. Interscience Publ., N.Y., 1959;
R.J.Glauber and G.Matthiae, {\sl Nucl. Phys.} {\bf B21} (1970) 135.

\bibitem{12} %%%%
W.Fabian and H.Arenh\"{o}vel, {\sl Nucl. Phys.} {\bf A 445}
(1976), 461.
H.Arenh\"{o}vel and M.Sanzone, {\sl Few Body Systems},
{\sl Supplementum 3} (1991).

\bibitem{13} %%%%
G.D.Alkhazov, S.I.Belostotsky, and A.A.Vorobyev, {\sl Phys. Rep.}
{\bf C42} (1978) 89.

\bibitem{14} %%%%
T.Lasinski et al., {\sl Nucl. Phys.} {\bf B37} (1972) 1.

\bibitem{15} %%%%
C.Lechanoine-LeLuc and F.Lehar, {\sl Rev. Mod. Phys.} {\bf 65}, 47
(1993).

\bibitem{16}
B.Jennings and B.Z.Kopeliovich, {\sl Phys. Rev. Lett.} {\bf 70}
(1993) 3384;
N.N.Nikolaev, A.Szczurek, J.Speth, J.Wambach, B.G.Zakharov,
and V.R.Zoller, {\sl Phys. Lett.} {\bf B317} (1993) 287;
A.Bianconi, S.Boffi, and D.E.Kharzeev, {\sl Phys.
Lett.} {\bf B325} (1994) 294.

\bibitem{17}
T.G.O'Neill et al., submitted to {\sl Phys. Rev. Lett.} (1994).

\bibitem{18} %%%%
C.Ciofi degli Atti and S.Simula, {\sl Phys. Lett.} {\bf B325}
(1994), 276.


\end{thebibliography}
\end{document}